\documentclass[prl,twocolumn]{revtex4}
\usepackage{graphics}

\begin{document}
\title{Adaptive homodyne measurement of optical phase}
\author{Michael A. Armen}
%\email{armen@caltech.edu}
\author{John K. Au}
\author{John K. Stockton}
\author{Andrew C. Doherty}
\author{Hideo Mabuchi}
\affiliation{Norman Bridge Laboratory of Physics 12-33, California
Institute of Technology, Pasadena, CA 91125}
\begin{abstract}
We present an experimental demonstration of the power of real-time
feedback in quantum metrology, confirming a theoretical prediction
by Wiseman \cite{Wise95} regarding the superior performance of an
adaptive homodyne technique for single-shot measurement of optical
phase. For phase measurements performed on weak coherent states
with no prior knowledge of the signal phase, we show that the
variance of adaptive homodyne estimation approaches closer to the
fundamental quantum uncertainty limit than any previously
demonstrated technique.  Our results underscore the importance of
real-time feedback for reaching quantum performance limits in
coherent telecommunication, precision measurement and information
processing.
\end{abstract}
\date{March 31, 2002}  \maketitle

Quantum mechanics complicates metrology in two complementary ways.
First it forces us to accept the existence of intrinsic
uncertainty in the value of observables such as the position,
momentum, and phase of an oscillator. Such quantum uncertainty
exists even when the {\em state} of the system being measured has
been prepared in a technically flawless way. Even the vacuum state
of a single-mode optical field, for example, exhibits `zero-point'
fluctuations in the amplitudes of its electric and magnetic
components. The same is true of the optical coherent states
representative of the output of an ideal laser.  These
uncertainties limit the sensitivity of ubiquitous measurement
techniques such as laser interferometry.  Squeezed states of light
are challenging to produce but have reduced uncertainties that can
in principal be exploited for improved sensitivity in optical
metrology \cite{Ref2}. To date, however, the achievable
performance gains have not been sufficient to motivate their use
in practical applications.

Quantum mechanics presents a second major obstacle to precision
measurement by making it generally quite difficult to realize
ideal {\em measurement procedures}, whose inherent inaccuracy is
small enough to reveal the intrinsic uncertainty limits associated
with quantum states.  Let us refer to such ideal measurements as
being Uncertainty-Limited (UL). Clearly, the implementation of UL
measurement procedures is essential for any application that seeks
to take advantage of exotic quantum states with reduced intrinsic
uncertainties. In connection with squeezed states of light,
measurements of optical quadrature amplitudes do constitute an
important class of UL measurements that actually can be
implemented in practice (via homodyne detection). But this is an
unusual case as UL measurement schemes have not previously been
demonstrated even for closely related observables such as optical
phase, despite intense historical interest in their quantum
properties \cite{Ref3,Ref4}.  This shortcoming is not merely one
of achievable signal-to-noise ratio in realistic experiments --
there has yet to be any demonstration of a measurement procedure
that is capable {\em even in principle} of achieving UL estimation
of true optical phase (as opposed to phase-quadrature amplitude),
for coherent or any other pure states of an optical field.

In this article we present an experimental demonstration of the
surprising efficacy of {\em real-time feedback} in the development
of UL measurement procedures. We do this in the context of
measuring the optical phases of weak pulses of light, following a
theoretical proposal by Wiseman \cite{Wise95}. This metrological
task can be motivated by a coherent optical communication scenario
in which information is encoded in the phase of laser pulses that
must travel long distances between the sender and receiver. In
such a context the receiver is likely to be faced with decoding
information carried by optical wave-packets whose quantum states
correspond to coherent states with low mean photon number (as a
result of optical attenuation). If the sender is encoding
information in an efficient manner, the variation of phase from
one optical wave-packet to the next should be uniformly
distributed over the entire interval from zero to $2\pi$.  Hence,
the receiver would ideally like to implement a {\em single-shot}
UL measurement procedure for estimating the phase of each
individual pulse \cite{Ref5}. The variance of such optimal phase
estimates should be limited only by the intrinsic quantum
uncertainty associated with optical coherent states of the given
mean photon number.

There is no known experimental procedure to accomplish this goal
exactly. Prior to Wiseman's proposal \cite{Wise95} it had been
widely believed \cite{Ref6} that the best feasible strategy for
single-shot phase decoding should be heterodyne detection, which
for coherent signal states can in principle achieve a measurement
variance that is only a factor of two greater than the intrinsic
uncertainty limit \cite{Ref7}. In what follows we will thus
consider heterodyne phase estimation as the benchmark we need to
surpass. It is important to compare the current measurement
scenario with what arises in applications such as the
implementation of optical frequency standards (optical clocks),
where optical phases must also be measured optimally but where
there is a guarantee that the phases vary only over a range much
less than $\pi/2$.  In such cases, as mentioned above, essentially
UL measurement can be achieved by employing fixed-quadrature
homodyne detection with a local oscillator whose phase is held at
an offset of $\pi/2$ radians from the expected signal phase.
Real-time feedback can thus be seen as a key ingredient in
formulating an UL scheme for measurement scenarios in which we
have no prior knowledge of the signal phase -- feedback enables
protocols in which the local oscillator phase is {\em adapted}, in
real time, to the phase of each individual signal pulse
\cite{Ref8}. As each optical pulse has some spatiotemporal extent,
the measurement signal generated by the leading edge of a given
pulse can be used to form a preliminary estimate of its phase,
which is used promptly to adjust the local oscillator settings to
be optimal for that pulse. (Note that this essentially amounts to
the implementation of a quantum-noise limited phase-lock loop; the
scheme relies on lock acquisition from the first one or two
photons of signal in each pulse.) Detailed theoretical analyses of
such schemes by Wiseman and co-workers have led to the striking
realization that, despite extremely low signal-to-noise ratio in
the feedback loop, adaptive homodyne measurement can be
essentially UL for optical pulses with mean photon number of order
ten (or greater), and benefits of real-time adaptation should
still be evident for mean photon numbers $\sim 1$.  We turn now to
our experimental test of these predictions.

The data plotted in Fig.~1A demonstrate the superiority of an
adaptive homodyne measurement procedure (`adaptive') to the
benchmark heterodyne measurement procedure (`heterodyne') for
making single-shot estimates of optical phase. As described in
detail below, we perform these measurements on optical pulses of
50 $\mu$s duration derived from an intensity-stabilized cw laser.
We have also plotted the theoretical prediction for the variance
of ideal heterodyne measurement, both with (thin solid line) and
without (dotted line) correction for a small amount of excess
electronic noise in the balanced photocurrent.  The excellent
agreement between the heterodyne data and theory indicates that we
have no excess phase noise (technically-ideal preparation of
coherent signal states) and validates electronic calibrations
involved in our data analysis \cite{Ref9}.  In the range of $\sim
10-300$ photons per pulse, most of the adaptive data lies below
the absolute theoretical limit for heterodyne measurement (dotted
line), and all of it lies below the curve that has been corrected
for excess electronic noise (which also has a detrimental effect
on the adaptive data). Quantitatively, we note that for $\approx
50$ photons per pulse the adaptive data point sits 6.5 standard
deviations below the absolute heterodyne limit (note logarithmic
scale).

For intermediate values of the photon number the adaptive
variances should ideally be even lower, but the performance of our
experiment is limited by finite feedback bandwidth. For signals
with large mean photon number, the adaptive scheme is inferior to
heterodyne because of excess technical noise in the feedback loop.
As the intrinsic phase uncertainty of coherent states becomes
large for very low photon numbers, the relative differences among
the expected variances for adaptive, heterodyne, and ideal
estimation become small.  Accordingly, we have been unable to beat
the heterodyne limit for phase-estimation {\em variances} with
adaptive homodyne measurement for mean photon numbers $N\alt 8$.
(Note that all theoretical curves in Fig.~1A correspond to
asymptotic expressions valid for large mean photon numbers;
corrections are small for $N\agt 10$.)  However, we are able to
show that the estimator {\em distribution} for adaptive homodyne
remains narrower than that for heterodyne detection even for
pulses with mean photon number down to $N\approx 0.8$. In Fig.~1B
we display the adaptive and heterodyne phase-estimator
distributions for a signal size corresponding to $\approx 2.5$
photons. Note that we have plotted the distributions on a
logarithmic scale, and that the mean phase has been subtracted off
so that the distributions are centered at zero. The horizontal
axis can thus be identified with estimation error. The adaptive
phase distribution has a smaller `Gaussian width' than the
heterodyne distribution, but exhibits rather high tails. The
observed shape of the adaptive distribution agrees qualitatively
with predictions \cite{Ref10} based on quantum estimation theory.
Theoretical analysis suggests that the high tails are a
consequence of changes in the sign of the photocurrent caused by
vacuum fluctuations of the optical field, which become comparable
to the coherent optical signal for $N\sim 1$. The feedback
algorithm responds to these photocurrent inversions by locking to
incorrect phase values with an error that tends towards $\pi$.

Accurately assessing the performance of a single-shot measurement
requires many repetitions of the measurement under controlled
conditions. Fig.~2A shows a schematic of our apparatus. Light from
a single-mode cw Nd:YAG laser is first stripped of excess
intensity noise by passage through a high-finesse Fabry-Perot
cavity (not pictured) with ringdown time $\approx 16$ $\mu$s; the
transmitted beam is shot-noise limited above $\sim 50$ kHz. This
light enters the Mach-Zehnder interferometer (MZI) at
beam-splitter 1 (BS1), creating two beams with well-defined
relative phase. The local oscillator (LO) is generated using an
acousto-optic modulator (AOM) driven by an RF synthesizer RF1 at
84.6 MHz, yielding $\sim 230$ $\mu$W of frequency-shifted light.
The signal beam corresponds to a frequency sideband created by an
electro-optic modulator (EOM) driven by an RF synthesizer (RF2)
that is phase locked to RF1. The power (5 fW to 5 pW) and pulse
length (50 $\mu$s) of the signal beam are controlled by changing
the amplitude of RF2 and by switching it on/off. A pair of
photodetectors collect the light emerging from the two output
ports of the final 50/50 beam-splitter (BS2); the difference of
their photocurrents (balanced photocurrent) provides the basic
signal used for either heterodyne or adaptive phase estimation. At
our typical LO power, the photodetectors supply 6 dB of shot noise
(over electronic noise) in the difference photocurrent from to
$\sim 1$ kHz to 10 MHz. We perform adaptive homodyne measurement
by feedback to the phase of RF2, which sets the (instantaneous)
relative phase between signal and LO \cite{Ref12}. Our feedback
bandwidth $\sim 1.5$ MHz is limited by the maximum slew rate of
RF2. Real-time electronic signal processing for the feedback
algorithm is performed by a Field Programmable Gate Array (FPGA)
that can execute complex computations with very low latency
\cite{Ref13}. Our feedback and phase estimation procedures
correspond to the `Mark II' scheme of Wiseman and co-workers
\cite{Ref10}, in which the photocurrent is integrated with
time-dependent gain to determine the instantaneous feedback
signal. The final phase estimate depends on the full history of
the photocurrent and feedback signal. For heterodyne measurements
we turn off the feedback to RF2 and detune it from RF1 by 1.8MHz;
phase estimates are made by the standard method of I/Q
demodulating the photocurrent beat note. For both types of
measurement we store the photocurrent $I(t)$ and feedback signal
$\Phi(t)$ on a computer for post-processing.

In Fig.~2B we show the photocurrent and feedback signal of three
consecutive adaptive phase measurements for $N\approx 8$. At the
beginning of each measurement, feedback of the photocurrent shot
noise causes the relative phase between the signal and LO to vary
randomly. As more of the pulse is detected, the information gained
is used to drive the signal-LO phase towards an optimal value.
Phase estimation variances are established using ensembles of
$\approx 150$ consecutive single-shot measurements. Although each
optical pulse is the subject of an {\em independent} single-shot
measurement, we fix the signal phase over the length of each
ensemble in order to accurately determine the estimation variance.
It is important to note that each data point in Fig.~1A
corresponds to an average of variance estimates from many
ensembles, each with a different (random) signal phase.

Ideally, the performance of a single-shot phase measurement
procedure should be independent of the signal phase.  In our
experiment, this implies that the measurement variances should be
independent of the initial relative phase between signal and local
oscillator.  In Fig.~1C we display a polar plot of the adaptive
variance (radial) versus initial signal phase (azimuth).  The data
were taken using signals with mean photon number $\approx 50$.
Each data point corresponds to the variance of an ensemble of
phase measurements taken at a fixed phase. The double solid lines
indicate the (one sigma) performance of our heterodyne
measurements. It is clear from this data that the performance of
the adaptive homodyne scheme is essentially independent of initial
phase, and is consistently superior to heterodyne measurement.

The photon number per pulse, $N$, is determined by extracting
optical amplitude information from the balanced photocurrent in
heterodyne mode. The typical experimental procedure is to fix the
signal amplitude, take an ensemble of heterodyne measurements
(which yields both heterodyne phase estimates and an estimate of
$N$), and then take an ensemble of adaptive homodyne measurements.
The photon number assigned to the subsequent ensemble of adaptive
measurements is $0.95N$, where the relative calibration factor
arises from the measured response of the EOM.

In conclusion, we have presented an experimental demonstration of
adaptive homodyne phase measurement.  For pulses with mean photon
number $\sim 10-300$ our measured variances approach closer to the
intrinsic phase-uncertainty limit of coherent states than any
previously demonstrated technique, including {\em ideal}
heterodyne detection. These results establish the technical
feasibility of broadband quantum-noise-limited feedback for
adaptive quantum measurement \cite{Ref14}, quantum feedback
control \cite{Ref15}, quantum error correction \cite{Ref16}, and
studies of conditional quantum dynamics \cite{Ref17,Ref18}.

\begin{acknowledgements}
This work was supported by the NSF (PHY-9987541, EIA-0086038) and
ONR (N00014-00-1-0479). JKS acknowledges the support of a Hertz
Fellowship.
\end{acknowledgements}

\vfill\eject

\begin{figure*}
\caption{Experimental results from the adaptive and heterodyne
measurements.  (a) Adaptive (blue circles) and Heterodyne (red
crosses) phase estimate variance vs. pulse photon number. The blue
dashed-dotted line is a second order fit to the adaptive data. The
thin lines are the theoretical curves for heterodyne detection
with (solid) and without (dotted) corrections for detector
electronic noise. The thick solid line denotes the fundamental
quantum uncertainty limit, given our overall photodetection
efficiency.  (b) Phase-estimator distributions for adaptive (blue
circles) and heterodyne (red crosses) measurements, for pulses
with mean photon number $\approx 2.5$. (c) Polar plot showing the
variance of adaptive phase estimates (blue dots) for different
signal phases (mean photon number $\approx 50$).  The solid blue
line is a linear fit to the data.  The double red lines indicate
the one sigma range for heterodyne measurements averaged over
initial phase.}
\end{figure*}

\begin{figure*}
\caption{(a) Apparatus used to perform both adaptive homodyne and
heterodyne measurements (see text). Solid lines denote optical
paths, and dashed lines denote electrical paths.  (b) Photocurrent
$I(t)$ (red above), and feedback signal $\Phi(t)$ (green below),
for three consecutive adaptive homodyne measurements with
$N\approx 8$.  The x-axis represents time for both signals. The
y-axis scale indicates the absolute phase shifts made by the
feedback signal. The photocurrent is plotted on an arbitrary
scale.}
\end{figure*}


\begin{thebibliography}{99}

\bibitem{Wise95} H.\ M.\ Wiseman, Phys.\ Rev.\ Lett.\ {\bf 75}, 4587 (1995).

\bibitem{Ref2} K.\ Schneider {\it et al.}, Opt.\ Express {\bf 2}, 59 (1998).

\bibitem{Ref3} See special issue of Phys.\ Scr., {\bf T48} (1993).

\bibitem{Ref4} U.\ Leonhardt {\it et al.}, Phys.\ Rev.\ A {\bf 51}, 84 (1995).

\bibitem{Ref5} Note that this precludes the use of techniques such as homodyne
tomography, which rely on multiple measurements on an ensemble of
identically-prepared systems.

\bibitem{Ref6} C.\ M.\ Caves, P.\ D.\ Drummond, Rev.\ Mod.\ Phys.\ {\bf 66}, 481 (1994).

\bibitem{Ref7} This factor of two derives from the fact that heterodyne
detection samples both the amplitude and phase quadrature
amplitudes, which are complementary observables.

\bibitem{Ref8} Y.\ Yamamoto {\it et al.}, in {\it Progress in Optics}, ed.\ E.\ Wolf
(North-Holland, Amsterdam, 1990), Vol.\ {\bf 28}, p. 150.

\bibitem{Ref9} Our absolute calibration of the horizontal axis has been
adjusted for overall photodetection efficiency $\eta$, which has
no bearing on the relative comparison of adaptive and heterodyne
measurement {\em procedures} but does affect our placement of the
fundamental uncertainty limit associated with coherent states.  We
have independently measured $\eta\approx 0.56$, including detector
quantum efficiency and homodyne fringe visibility, and have
indicated the fundamental limit for phase estimation given this
value of $\eta$. For phase measurements performed on squeezed
states $\eta$ would need to be optimized, but its value and the
corresponding correction to the photon-number calibration are not
of direct concern in our current investigation involving coherent
states.

\bibitem{Ref10} H.\ M.\ Wiseman and R.\ B.\ Killip, Phys.\ Rev.\ A {\bf 57}, 2169 (1998).

\bibitem{Ref12} In principle one can modulate either the signal phase or the
LO phase; we do the former to avoid interference between the
adaptive homodyne feedback loop and an auxiliary feedback loop
used to stabilize the MZI.

\bibitem{Ref13} J.\ K.\ Stockton, M.\ Armen, and H.\ Mabuchi, in preparation
(quant-ph/0203143 at http://www.arXiv.org).

\bibitem{Ref14} D.\ W.\ Berry, H.\ M.\ Wiseman, Phys.\ Rev.\ A {\bf 6301}, 013813 (2001).

\bibitem{Ref15} A.\ C.\ Doherty {\it et al.}, Phys.\ Rev.\ A {\bf 6201}, 012105 (2000).

\bibitem{Ref16} C.\ Ahn, A.\ C.\ Doherty, and A.\ Landahl, Phys.\ Rev.\ A, {\bf 65}, 042301 (2002).

\bibitem{Ref17} W.\ P.\ Smith {\it et al.}, in preparation
(quant-ph/0202063 at http://www.arXiv.org).

\bibitem{Ref18} H.\ Mabuchi and H.\ M.\ Wiseman, Phys.\ Rev.\ Lett.\ {\bf 81}, 4620 (1998).

\end{thebibliography}
\end{document}